\documentclass[prb,twocolumn,showpacs,floatfix]{revtex4-1}

\usepackage[utf8]{inputenc}
\usepackage{graphicx}
\usepackage{flafter}
\usepackage{amsmath}
\usepackage{amssymb}
\usepackage{float}
\usepackage{color}
\usepackage{subfigure}
\definecolor{dg}{rgb}{0.00, 0.40, 0.29}

\let \k \relax
\newcommand{\k}{{\bf k}}
\newcommand{\p}{{\bf p}}
\newcommand{\q}{{\bf q}}

\newcommand{\pdagger}{{\phantom{\dagger}}}

\begin{document}

%\title{Excitonic Josephson effect in coupled double-layer graphene systems}
\title{Excitonic Josephson effect in double-layer graphene junctions}

\author{B. Zenker$^1$, H. Fehske$^1$, and H. Beck$^2$}
\affiliation{$^1$Institut f{\"u}r Physik,
                  Ernst-Moritz-Arndt-Universit{\"a}t Greifswald,
                  D-17489 Greifswald, Germany \\
                  $^2$D{\'e}partement de Physique and Fribourg Center for Nanomaterials, 
                  Universit{\'e} de Fribourg,
                  CH-1700 Fribourg, Switzerland}
\date{\today}
\begin{abstract}
We show that double-layer graphene (DLG), where an external potential induces a charge-imbalance between $n$- and $p$-type layers, is a promising candidate to realize an exciton condensate in equilibrium.  To prove this phenomenon experimentally, we suggest coupling two  DLG systems, separated by a thin insulating barrier, and measuring the excitonic Josephson effect.  For this purpose we calculate the ac and dc Josephson currents induced by tunneling excitons and show that the former only occurs when the gate potentials of the DLG systems differ, irrespective of the phase relationship of their excitonic order parameters. A dc Josephson current  develops if a finite order-parameter phase difference exists  
between two coupled DLG systems with identical gate potentials.  
\end{abstract}
\maketitle

The search for the long ago predicted excitonic insulator (EI) state has recently stimulated a lot of experimental work, e.g., on pressure sensitive rare-earth chalgogenides, transition metal dichalcogenides or tantalum chalcogenides~\cite{BSW91,WBM04,CMCBDGBAPBF07,WSTMANTKNT09,MMAB12b}.  Theoretically the excitonic instability is expected to happen, 
when semimetals with very small band overlap or semiconductors with very small band gap are cooled to very low temperatures~\cite{Mo61,Kno63}.
To date there exists no free of doubt realization of the EI, however, and even the applicability of the original EI scenario to the above material classes is a  controversial issue~\cite{WNS10,MMAB12b,ZFBMB13,KTKO13}. There are serious arguments why the EI in these bulk materials, if present at all, resembles rather a charge-density-wave state than a ``true'' superfluid exciton condensate exhibiting off-diagonal long-range order~\cite{HR68a,ZFB14}.

On these grounds  a non-ambiguous experimental proof of a macroscopic phase coherent exciton condensate would be highly desirable.
Spectroscopic analyses have not established an exciton condensate  so far. The characteristics of junction devices, where at least in one component an EI is realized, may lead to valuable insights in this respect~\cite{RS14}. Due to the proximity effect a high resistance should appear across a semimetal-EI junction that distinctly differs  from that of a semimetal-semiconductor device~\cite{RS05}. In coupled quantum wells,  Josephson oscillations should accompany exciton condensation~\cite{RS09,SSPM08}. Here we will pursue a similar idea, namely that a Josephson-type 
tunnel current might appear when two EI systems are coupled to each other by a thin insulating barrier such that coherence is established between the condensates.

Two-layer systems of spatially separated electrons and holes that feature an attractive interlayer electron-hole coupling are particularly suitable  for a Josephson-type tunnel experiment. In this  case a condensate of  excitons might occur when the tunneling between the layers is negligible, but the corresponding Coulomb interaction is not~\cite{BJL04}. Double-layer systems  thereby inhibit the obstacles coming from interband transitions or the coupling to  phonons, which inevitably occur in bulk materials and prevent a possible exciton condensation by destroying the $U(1)$ symmetry~\cite{LY76a,LZ96,ZFB14}.  It is also advantageous that (exciton) tunneling effects are experimentally well accessible in double-layer systems. 
\begin{figure}[h]
\centering
\includegraphics[width=0.8\linewidth]{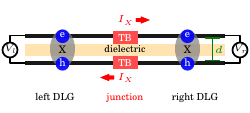} \\
\begin{minipage}{0.19\linewidth}
\includegraphics[width=\linewidth]{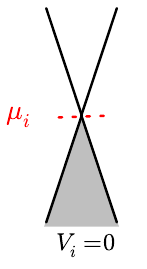}
\end{minipage}
\begin{minipage}{0.38\linewidth}
\includegraphics[width=\linewidth]{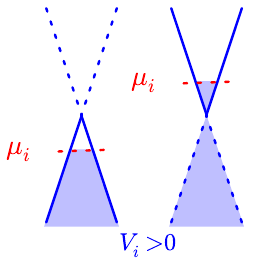}
\end{minipage}
\hspace{\fill}
\begin{minipage}{0.38\linewidth}
\includegraphics[width=\linewidth]{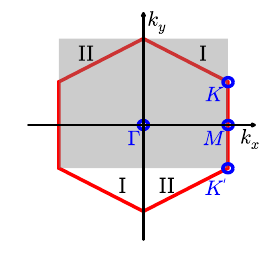}
\end{minipage}
\caption{(Color online) Upper panel: Schematic setup to verify an excitonic Josephson effect. 
Two DLG systems are separated by a thin tunnel barrier (TB).  By external potentials $V_l$ and $V_r$  the chemical potential (particle numbers) 
can be tuned in each layer of the left and right DLG, respectively.  Here the charge carriers are electrons ($e$) and holes ($h$) in the 
upper (lower) $n$-type ($p$-type) layer. The layers are separated by a dielectric of thickness $d$. Tunneling of coherent electron-hole pairs (excitons) will induce an electron current $I_X$ through the barrier. Obviously, the current in the lower layer equals the current in the 
upper layers in modulus but flows in the opposite direction.
Lower panel: Band structure near the $K$ point and chemical potential $\mu_i$ of neutral DLG with $V_i=0$ (left) and gated DLG where $V_i>0$  (middle). First Brillouin zone of DLG (red hexagon) with high-symmetry points (right). Backfolding parts I and II, the grey rectangle  shows 
the  representation of the Brillouin zone we use in Fig.~\ref{fig:OPs}.}
\label{fig:setup}
\end{figure}

Graphene-based double layers (separated by an adequate dielectric, e.g., hexagonal boron nitride or SiO$_2$) show great promise for realizing  
a corresponding setup (see Fig.~\ref{fig:setup}). For double-layer graphene (DLG), a gate-bias across the  layers creates a charge imbalance, 
whereupon the attractive Coulomb interaction between the excess electrons and holes on opposite layers raises the possibility of exciton formation.
While a fine tuning of the band gap can be achieved by the external potential, the recombination of electrons and holes 
can be fully suppressed by the dielectric~\cite{GGKNTGMMWTP12}. Then, in the weak coupling regime, 
exciton condensation is triggered by a Cooper-type instability, where the particle-hole symmetry of the 
system ensures a perfect nesting between the electron Fermi surface and its hole counterpart in the $n$- and $p$-type layers of a biased DLG system~\cite{MBSM08,NL10,HZW11}.   Placing an (insulating) tunnel barrier between two such DLG systems, the Josephson current can be used to analyze whether or not  an  exciton condensate has been formed in each of the subsystems.  We note that a similar setup was recently proposed to study thermal transport in a temperature-biased exciton-condensate junction~\cite{ZJ13}.  

The tight-binding Hamiltonian we assume for a DLG subsystem $i$ ($i=l,r$, left or right) has the form 
\begin{eqnarray}
H_{i} &=& \sum_{\k} \varepsilon_{\k i}^+ a_{\k i}^\dagger a_{\k i}^\pdagger
+ \sum_{\k} \varepsilon_{\k i}^- b_{\k i}^\dagger b_{\k i}^\pdagger
\nonumber \\
&&+\frac{1}{N} \sum_{\k,\k',\q} U_{\k,\k',\q} \, a_{\k+\q i}^\dagger 
a_{\k i}^\pdagger b_{\k'-\q i}^\dagger b_{\k' i}^\pdagger  \,,
\label{H_i_undec} 
\end{eqnarray}
where $a_{\k i}^{(\dagger)}$ and $b_{\k i}^{(\dagger)}$ annihilate (create) electron quasiparticles 
  in the $n$ layer and $p$ layer, respectively,  with in-plane momenta ${\k}$ and band dispersions 
\begin{eqnarray}
\varepsilon_{\k i}^{\pm} &=&\pm \gamma_0 
\left[ 3+2\cos\left(\sqrt{3} k_y\right) \right.
\nonumber \\
&&\left. +4\cos\left(\sqrt{3} k_y /2\right) 
\cos\left(3k_x/2\right) \right]^{\frac{1}{2}} \mp \mu_i \,.
\label{disp}
\end{eqnarray}
In the low carrier density regime, the effective band structure~\eqref{disp} should account for the effects of the 
intralayer Coulomb interaction. The corresponding particle transfer amplitude is parametrized by  $\gamma_0\simeq 2.8$ eV  (which defines the unit of energy in what follows) and the momenta $k_{x,y}$ by $a^{-1}$, where $a\simeq 0.142$ nm is the carbon-carbon distance within graphene's honeycomb structure (see Ref.~\onlinecite{CGPNG09}). 
  The external potential $V_i$ determines the chemical potential: $\mu_i=V_i /2$.
The interlayer Coulomb interaction leading to exciton formation is 
 \begin{equation}
U_{\k,\k',\q} = \kappa \frac{e^{-d|\q|}}{|\q|} \cos\left(\frac{\Phi}{2} \right)
\cos\left(\frac{\Phi'}{2} \right)\,,
\label{CoulombInt}
\end{equation}
where $\kappa = g_s 2\pi/\epsilon$, $g_s=2$, $\epsilon$ denotes the dielectric constant of the dielectric, and $N$ gives the total number of particles~\cite{ZJ08,PF12}. Since electron-hole recombination is prevented by the dielectric,  we neglected in Eq.~\eqref{H_i_undec} all interlayer Coulomb interaction terms that do not preserve the number  of electrons (or holes) in a single layer, e.g., $H_U \propto a_{\k+\q i}^\dagger a_{\k i}^\pdagger a_{\k'-\q i}^\dagger b_{\k' i}^\pdagger$ or $H_U \propto a_{\k+\q i}^\dagger b_{\k i}^\pdagger a_{\k'-\q i}^\dagger b_{\k' i}^\pdagger$.
Note that the model~\eqref{H_i_undec} exhibits a $U(1)$ symmetry, which causes the phase of the EI order parameter to be undetermined.

A mean-field decoupling of the Coulomb interaction yields
\begin{eqnarray}
\bar H_i &=& \sum_{\k} \varepsilon_{\k i}^+ a_{\k i}^\dagger a_{\k i}^\pdagger
+ \sum_{\k} \varepsilon_{\k i}^- b_{\k i}^\dagger b_{\k i}^\pdagger
\nonumber \\
&&+ \sum_\k \Delta_{\k i}^\ast b_{\k i}^\dagger a_{\k i}^\pdagger
+\sum_\k \Delta_{\k i} a_{\k i}^\dagger b_{\k i}^\pdagger ,
\label{H_i}
\end{eqnarray}
where we have introduced the EI order-parameter function
\begin{equation}
\Delta_{\k i}^\ast = -\frac{\kappa}{N} \sum_\q \frac{e^{-d|\q|}}{|\q|} 
\frac{1+\cos(\Phi)}{2} \langle a_{\k+\q i}^\dagger b_{\k+\q i}^\pdagger
\rangle \,,
\label{EIOP}
\end{equation}
where $\Phi=\Theta_{\k+\q}-\Theta_\k$ ($\Phi'=\Theta_{\k'+\q}-\Theta_{\k'}$) with $\Theta_\k=\arctan(k_y/k_x)$.
 The phase of $\Delta_{\k i}^\ast$ determines the phase of the ground-state wave function.

In view of the specifics of the graphene spectrum, which are suggestive of large screening effects, the correct approximation for the screening of the interlayer Coulomb interaction has been a controversial issue~\cite{KE08,LMS08,NPH14}.  A mean-field inclusion in the normal phase certainly overestimates screening and reduces $|\Delta_\k|$ in an unrealistic way~\cite{NPH14}. In addition, correlations between electrons and holes may substantially weaken the screening~\cite{LOS12}. As a result double-layer graphene-based systems are not that disadvantageous for the realization of an excitonic condensate as naively might be expected. Determining the transition temperature of the EI phase necessitates,  of course, an appropriate treatment of (dynamical) screening in the condensed phase~\cite{MBSM08, LOS12, NPH14}. This is beyond the scope of this work. In order to discuss the excitonic Josephson effect in the case of a realized EI ground state at zero temperature, Eqs.~\eqref{H_i_undec}-\eqref{EIOP} are adequate to leading order. 
 
\begin{figure}[h]
\centering
\begin{minipage}{0.49\linewidth}
(a) \\
\includegraphics[width=0.9\linewidth]{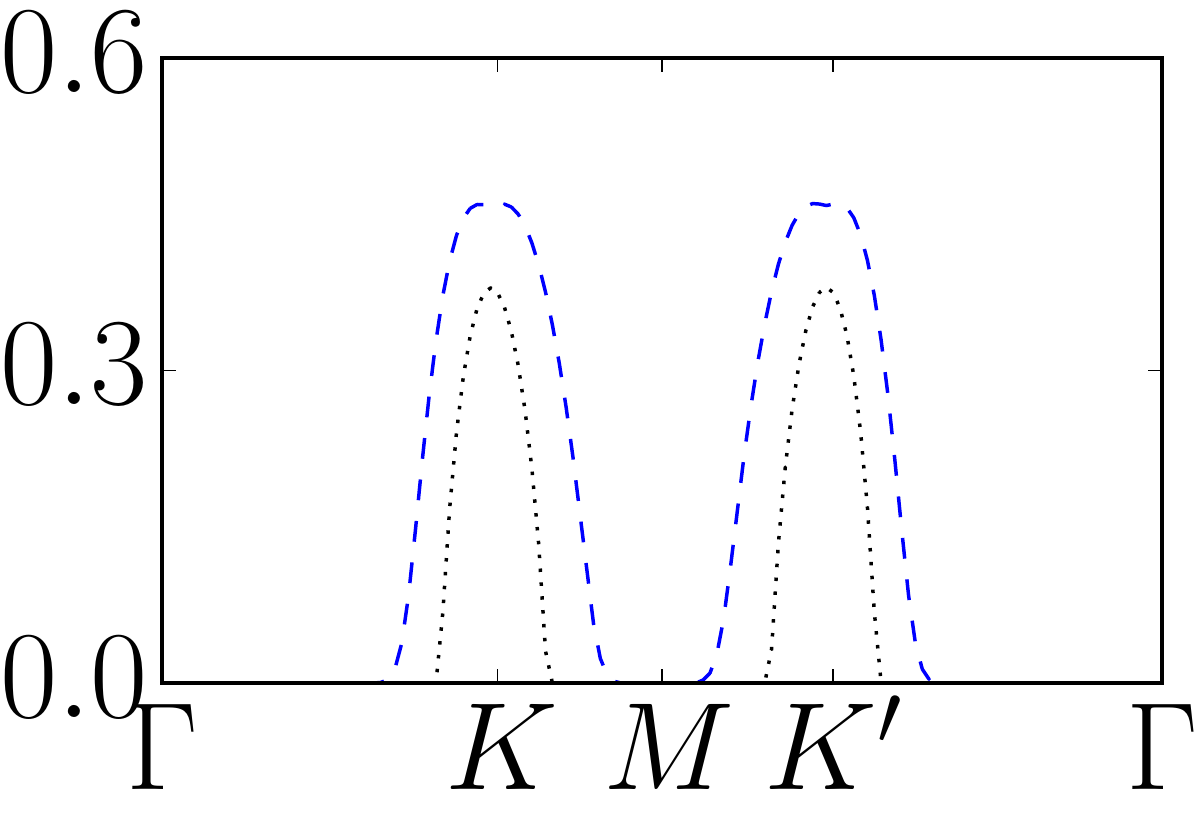}
\end{minipage}
\begin{minipage}{0.49\linewidth}
(b) \\
\includegraphics[width=0.9\linewidth]{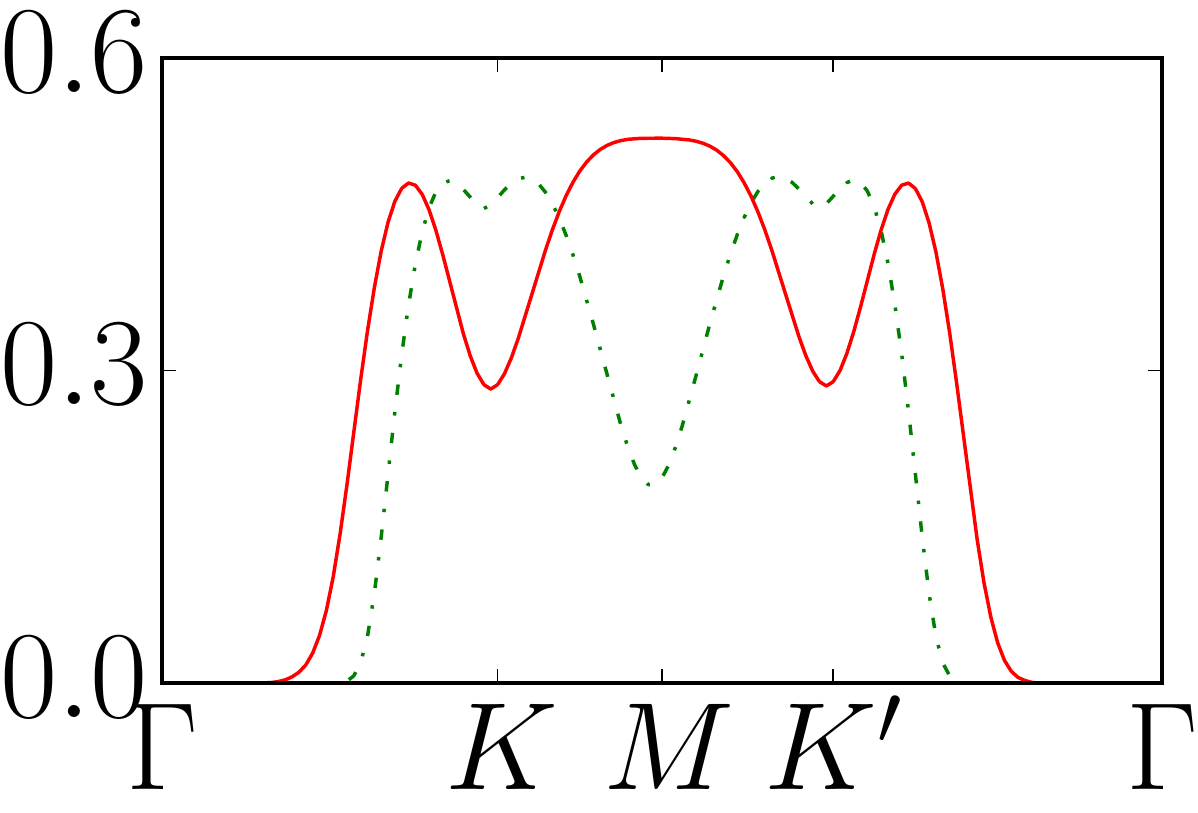}
\end{minipage}
\begin{minipage}{0.03\linewidth}
\hspace{\fill}
\end{minipage}
\begin{minipage}{0.36\linewidth}
(c) \\
\includegraphics[width=\linewidth]{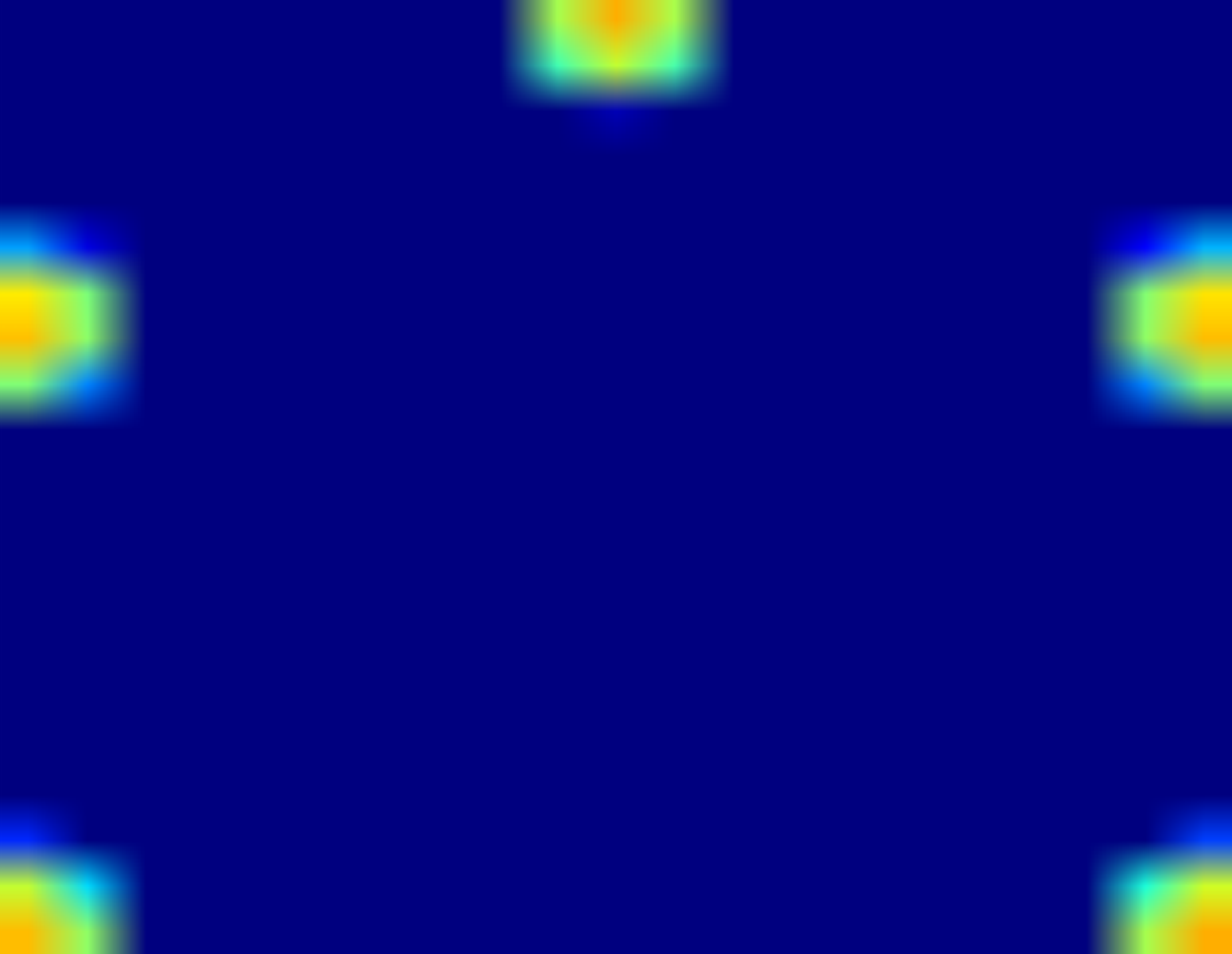}
\end{minipage}
\begin{minipage}{0.03\linewidth}
\hspace{\fill}
\end{minipage}
\begin{minipage}{0.1\linewidth}
\textcolor{white}{a}\\
\includegraphics[width=.9\linewidth]{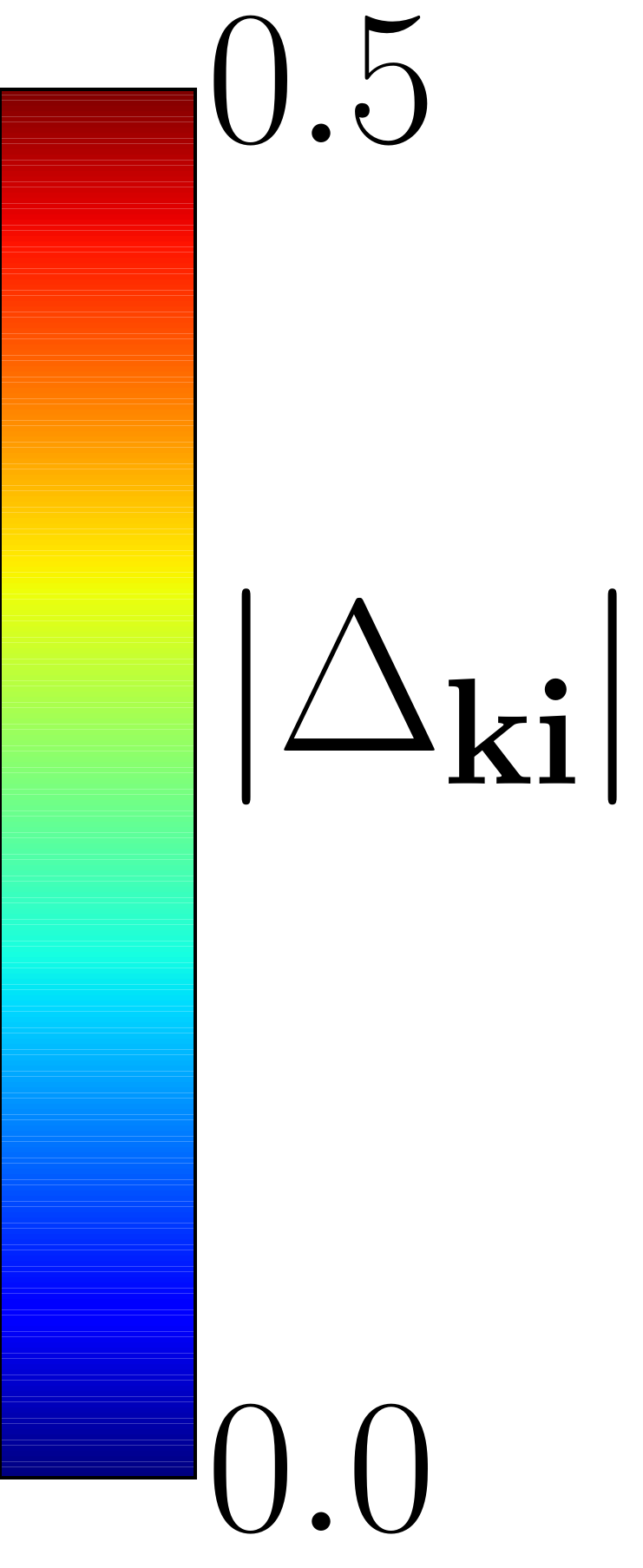}
\end{minipage}
\begin{minipage}{0.36\linewidth}
(d)\\
\includegraphics[width=\linewidth]{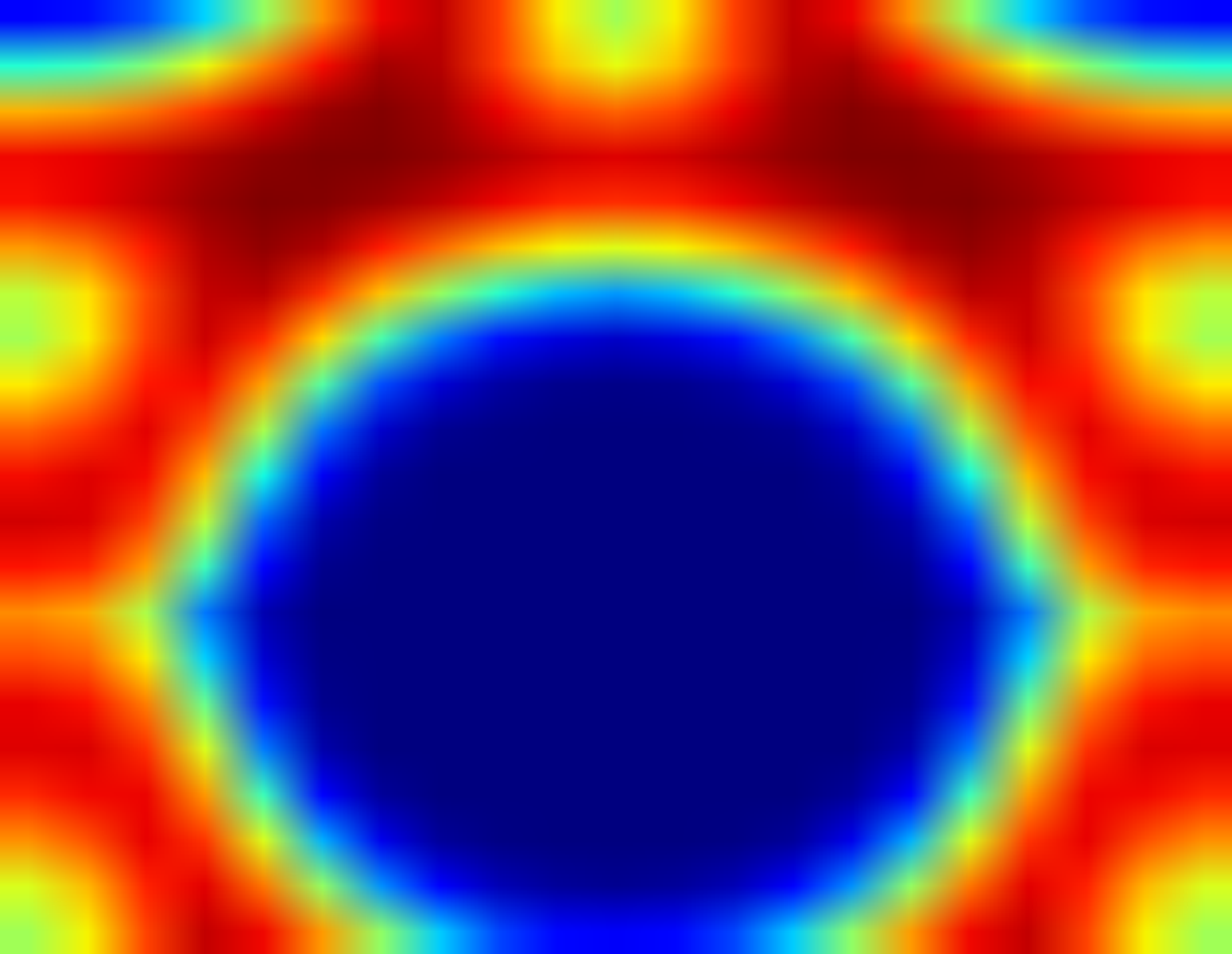}
\end{minipage}
\caption{(Color online) Modulus of the EI order parameter function for DLG in reciprocal space, $|\Delta_{\k i}|$,   
at different gate voltages $V_i$. The upper line plots give $|\Delta_{\k i}|$ along the high-symmetry directions of the Brillouin zone  (cf.~Fig.~\ref{fig:setup}),  for (a) $V_i=0.0$ (black dotted line), $V_i=0.5$ (blue dashed line) and (b) $V_i=1.0$ (green dot-dashed line), $V_i=2.0$ (red solid line).  
Corresponding intensity plots in the first Brillouin zone (using a $20\times 20$ grid in ${\bf k}$-space), for (c) $V_i=0.0$ and (d) $V_i=2.0$.} 
\label{fig:OPs}
\end{figure}

 We first solve the self-consistency equations for the EI order parameter at zero temperature, assuming $\kappa=7.0$ and  $d=2.5$. 
Figure~\ref{fig:OPs} shows its modulus  $|\Delta_{\k i}|$ within the first Brillouin zone. 
Apparently only electrons and holes near the Brillouin zone's $K$ or $K'$ points are bound into excitons. These particles occupy the states closest to the Fermi energy and therefore will be most susceptible for electron-hole pairing. Since the external potential fixes the position of  the Fermi energy, it determines the behavior of $\Delta_{\k i}$ as well. When $V_i$ is raised from zero, more and more states become available for an electron-hole pairing. As a result the EI order parameter function is finite in a larger region of the Brillouin zone and the total (momentum accumulated) order parameter increases. At $V_i\approx 2$ the EI order parameter attains its maximal value and starts to decrease if  the gate voltage gets larger until the EI phase breaks down at about $V_i\approx 6$, where the Fermi energy coincides with the upper (lower) edge of the conduction (valence) band. The relatively sharp boundary confining the area of bound electron-hole pairs indicates a BCS-like pairing~\cite{PBF10}.

A Josephson effect occurs when two DLG systems, where quantum coherence is realized, will be coupled to each other~\cite{Jo62}. 
Assuming that the ground states in both subsystems are described by macroscopic wave functions, $\Psi_i = |\Psi_i|e^{i\phi_i}$ 
($i=l,r$ index the left, respectively, right, subsystem),  which---for simplicity reasons---are assumed to be equal in modulus but may have a different phases.
The current density that flows between the left and right subsystem is 
$j = \frac{ie}{2}\left(\Psi_l \nabla \Psi_r^\ast - \Psi_r \nabla \Psi_l^\ast\right)$ 
(we set both $\hbar=1$ and the electron mass $m_e=1$; $e$ denotes the electron charge).
Obviously any finite phase difference $\Delta\phi=\phi_r-\phi_l$ prevents the current density  from vanishing, 
i.e.,  a persistent tunnel current flows through the barrier. 

The Hamiltonian of the coupled system is $H = \bar H_l + \bar H_r + H_{T}$, with 
\begin{eqnarray}
H_{T} &=& \sum_{\k,\p} T_{\k,\p} 
\left( a_{\k l}^\dagger a_{\p r}^\pdagger 
+ b_{\k l}^\dagger b_{\p r}^\pdagger \right)
\nonumber \\
&&+ \sum_{\k,\p} T_{\k,\p}^\ast \left(a_{\p r}^\dagger a_{\k l}^\pdagger 
+ b_{\p r}^\dagger b_{\k l}^\pdagger\right) ,
\label{H_T}
\end{eqnarray}
describing the tunnel process. 

Tunneling excitons cause an electron current in the $n$ layer. This current equals the one in the $p$ layer---which flows in the opposite direction, however---in modulus.
Below we adapt the approach outlined in Ref.~\onlinecite{Mah00} to the (coherent) exciton tunnel processes in DLG. 
For this we define the tunnel current as the time derivative of the number of electrons in the upper layer and use an $S$-matrix expansion approach~\cite{Mah00},
\begin{equation}
I(t) = e\langle \dot N_{a l}(t) \rangle = -ie \int_{-\infty}^t dt'
\left\langle \left[ \dot N_{a l}(t), H_{T}(t') \right] \right\rangle\,, 
\end{equation}
where $N_{a l} = \sum_\k a_{\k l}^\dagger a_{\k l}^\pdagger$, $\dot N_{a l}
=-i \sum_{\k,\p} T_{\k,\p} a_{\k l}^\dagger a_{\p r}^\pdagger 
+ i \sum_{\k,\p} T_{\k,\p}^\ast a_{\p r}^\dagger a_{\k l}^\pdagger$,   
 $\dot N_{a l}(t) = e^{iH't} \dot N_{a l} e^{-iH't}$,  and $H_{T}(t') = e^{iH' t'} H_{T} e^{-i H't'}$,  with $H'=\bar H_l + \bar H_r$.
Note that the chemical potential in the left and right DLG systems may differ. 
We introduce the applied junction voltage 
\begin{equation}
W = \mu_{r} - \mu_{l} ,
\label{W}
\end{equation}
and the operators $A(t) = \sum_{\k,\p} T_{\k,\p}^\ast a_{\p r}^\dagger(t) a_{\k l}^\pdagger(t)$  and $B(t) = \sum_{\k,\p} T_{\k,\p}^\ast b_{\p r}^\dagger(t) b_{\k l}^\pdagger(t)$.
%\begin{eqnarray}
%A(t) = \sum_{\k,\p} T_{\k,\p}^\ast a_{\p r}^\dagger(t) a_{\k l}^\pdagger(t) , \label{A}
%\\
%B(t) = \sum_{\k,\p} T_{\k,\p}^\ast b_{\p r}^\dagger(t) b_{\k l}^\pdagger(t) . \label{B}
%\end{eqnarray}
Although the specific choice of the tunnel matrix element will affect the Josephson current, we  leave the investigation of this quantitative facet for future work. Instead we focus on the influence of different chemical potentials and phases of the EI order parameters in the DLG systems,
which will dominate the physics.  Analyzing Josephson tunneling for normal superconductors  the combination $T_{\k, \p}T^*_{\k, \p}$ is usually approximated by $|T|^2$ times a phase factor, where both quantities are assumed to be independent of ${\bf k}$ and $\bf p$~\cite{Mah00}. In our case, the order parameter function has a rather complex  
momentum dependence, however (see Fig.~\ref{fig:OPs}). To avoid a fourfold numerical integration over the Brillouin zone, we employ  $T_{\k, \p}=\delta_{\k,\p}$, which---in the sense of a ``principle of proof'' calculation---will underestimate the effect we are looking for. Any more sophisticated treatment of the tunnel barrier will increase both the tunnel and normal currents. 
 
With that, the tunnel current takes the form 
\begin{align}
I&(t) = -(i)^2 e \int_{-\infty}^t dt' \left( 
\left\langle [A(t), A^\dagger(t')]\right\rangle e^{iW(t-t')}  \right.
\nonumber \\
&+ \left\langle [A(t), B^\dagger(t')]\right\rangle e^{iW(t+t')} 
+ \left\langle A(t),A(t')] \right\rangle e^{iW(t+t')}
\nonumber \\
&+ \left\langle [A(t),B(t')]\right\rangle e^{iW(t-t')}
- \left\langle [A^\dagger(t), A^\dagger(t')] \right\rangle e^{-iW(t+t')}
\nonumber \\
&- \left\langle [A^\dagger(t),B^\dagger(t')]\right\rangle e^{-iW(t-t')}
- \left\langle [A^\dagger(t),A(t')]\right\rangle e^{-iW(t-t')}
\nonumber \\
&\left.- \left\langle [A^\dagger(t), B(t')]\right\rangle e^{-iW(t+t')} \right).
\label{tot_current}
\end{align}

Let us first focus on the current due to tunneling excitons exclusively. This current is governed by the second term and eighth term of Eq.~\eqref{tot_current}:
\begin{equation}
I_X(t) = 2e\, {\rm Im}\left[ e^{2iWt} X_{\rm ret}(-W) \right]\,, 
\label{Xcurrent}
\end{equation}
containing the retarded Green's function 
$X_{\rm ret}(W) = \int_{-\infty}^\infty dt \, e^{iW(t-t')} X_{\rm ret}(t-t')$, 
$X_{\rm ret} (t-t') = -i\Theta(t-t') 
\left\langle [A(t), B^\dagger(t')] \right\rangle$.

Normal tunneling electrons are described by the first and the seventh term of Eq.~\eqref{tot_current}:
\begin{equation}
I_n(t) = -2e\, {\rm Im} Y_{\rm ret}(W)\,,
\label{ncurrent}
\end{equation}
where  
$Y_{\rm ret}(W) = \int_{-\infty}^\infty dt \, e^{iW(t-t')} Y_{\rm ret}(t-t')$,
$Y_{\rm ret} (t-t') = -i\Theta(t-t') 
\left\langle [A(t), A^\dagger(t')] \right\rangle$. 
It is worth noting that splitting up the current into a  quasiparticle (normal) and interference (oscillating Josephson) current was also carried out analyzing the thermal transport through Josephson junctions based on DLG~\cite{ZJ13}.

We now factorize $X_{\rm ret}(W)$ and $Y_{\rm ret}(W)$ into contributions stemming from the left and right subsystems. 
For these we use the mean-field Green's functions calculated with the Hamiltonian~\eqref{H_i}, which are expressed in terms of the quasiparticle states.

\begin{figure}[t]
\centering
\includegraphics[width=.85\linewidth]{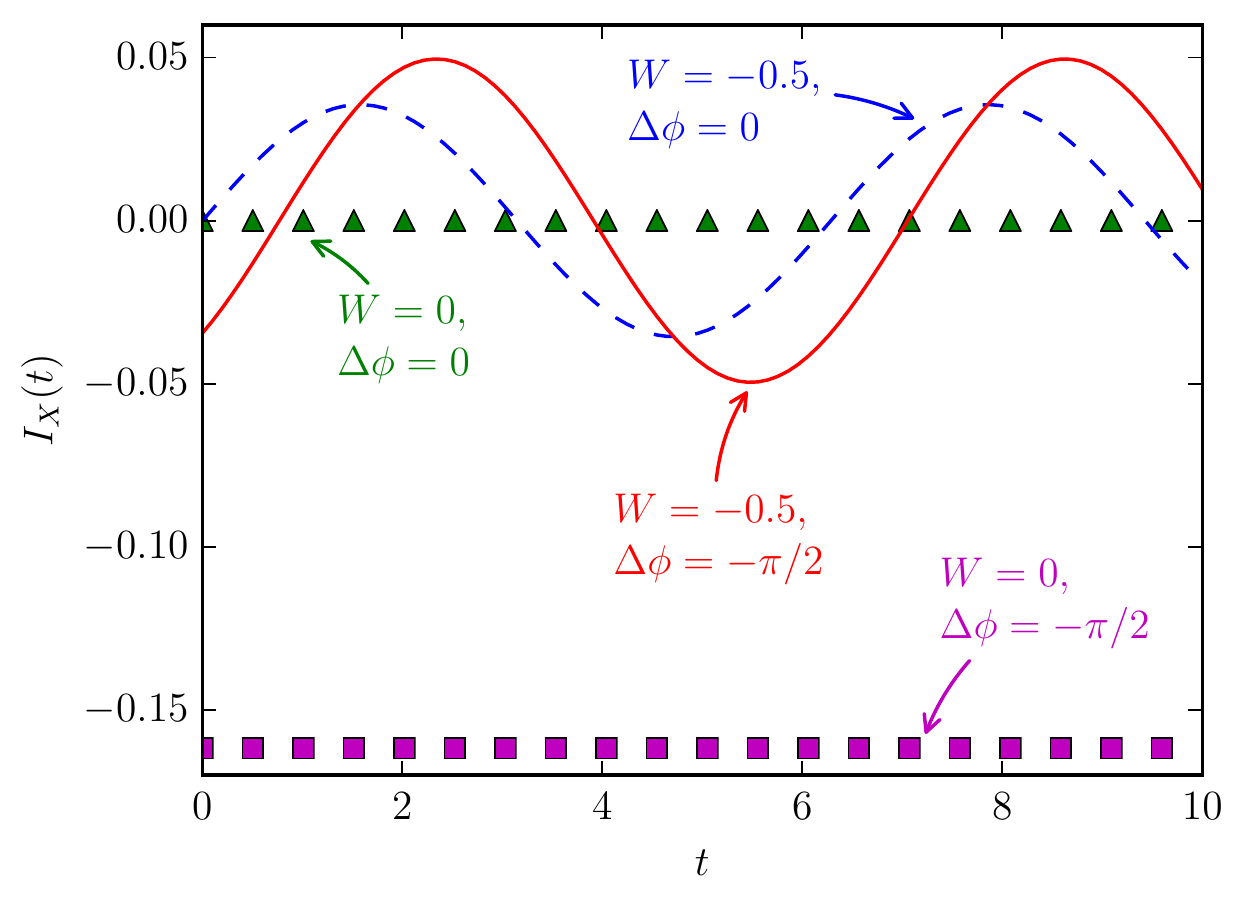}
\caption{(Color online) Excitonic Josephson current in case of (i) two identical DLG subsystems (green triangles, $V_l=1.0$ and $V_r=1.0$), (ii) equal left and right external potentials but different phases of the EI order parameters (magenta squares, $V_l=1.0$ and $V_r=1.0$),  (iii) different external potentials with EI order parameters having the same phase in the left and right DLG systems (blue dashed line, $V_l=0.0$ and $V_r=1.0$), 
and (iv) different phases and different external potentials  in both DLG subsystems (red continuous line, $V_l=0.0$ and $V_r=1.0$).}
\label{fig:currents}
\end{figure}

Figure~\ref{fig:currents} gives the time dependence of the tunnel current for four characteristic situations.
We first consider the case that two identical DLG systems are coupled by a thin barrier, i.e., $V_l=V_r$, $W=0$. Clearly, if both EI order parameters have the same phase, no current flows through the junction. A dc current arises when the left and the right systems differ in terms of the phase of the EI order parameter: $\Delta\phi=\phi_r-\phi_l=-\pi/2$ in Fig.~\ref{fig:currents}. An ac current appears, on the other hand, if a finite (constant) voltage is applied across the junction. Its frequency is $2W$, which coincides with the frequency of the Josephson current for coupled superconductors. An additional phase difference amplifies the tunnel-current amplitude and leads to a phase shift in the current.

\begin{figure}[t]
\centering
\includegraphics[width=.85\linewidth]{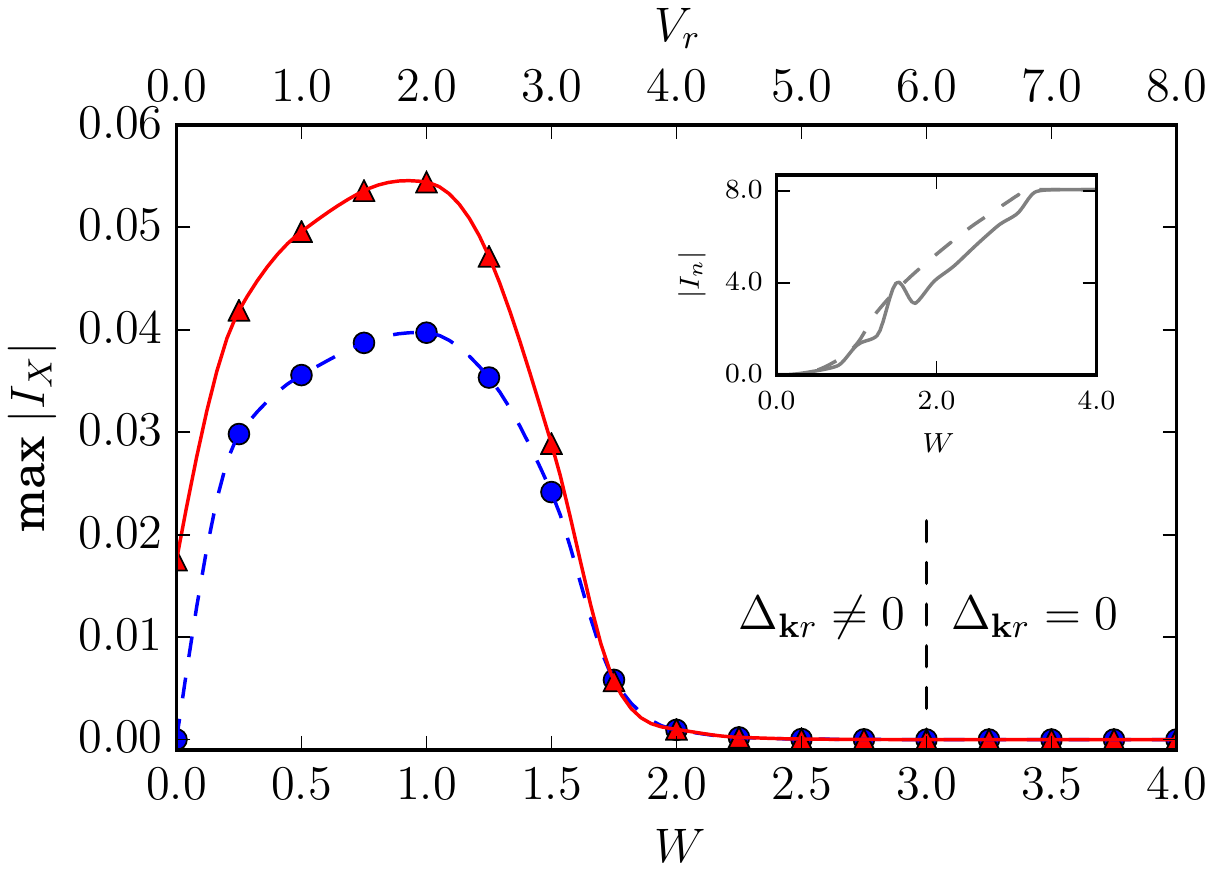}
\caption{(Color online) Amplitude of the excitonic Josephson current in dependence on the applied junction voltage in the case that the phases of the order parameters are equal ($\Delta\phi=0$, blue circles) or differ ($\Delta\phi=-\pi/2$, red triangles). The dashed and solid lines are interpolations to guide the eye. 
The inset shows the normal tunneling of electrons. The solid line considers $\Delta_{\k r}$ and, for comparison, the dashed line shows the current, when $\Delta_{\k r}$ is set to zero.}
\label{fig:IW}
\end{figure}

We finally  analyze the voltage-current characteristics as to its phase dependence. For this purpose, we switch off the gate voltage in the left DLG system ($V_{l}=0$) and fix its EI phase to $\pi$. Now the gate voltage in the  right DLG subsystem is tuned from $V_r=0$ to $V_r=8$ 
for two choices  of the right EI's phase: $\phi_r=\pi$ and $\phi_r=\pi/2$. The corresponding results are displayed in Fig.~\ref{fig:IW}.

Most notably, at $W=0$, a finite dc Josephson current  only appears if $\Delta\phi\neq 0$. Otherwise $\Delta\phi=0$ and $\Delta\phi\neq 0$ basically cause the same qualitative behavior. The amplitude increases with increasing voltage until it reaches its maximum at a junction voltage $|W|=1.0$. This coincides with the point, $V_r=2.0$, at which the EI order parameter in the right DLG system attains its largest value. If the voltages grows further the amplitude of the exciting Josephson current vanishes rapidly. Unfortunately the current becomes extremely small just before the EI phase completely breaks down. Therefore, the Josephson tunnel current is not very suitable for the precise determination of the EI phase boundary.
Figure~\ref{fig:IW} corroborates that a finite phase difference yields a larger current amplitude.

Note that only for small gate voltages ($W\lesssim0.4$) the amplitude of the excitonic Josephson tunnel current will be of the same order of magnitude as the amplitude of a current due to tunneling excitons. For higher voltages the normal tunnel current is up to two orders of magnitude 
larger (see the inset in Fig.~\ref{fig:IW}). That is why we propose an experimental setup, where the chemical potentials of the two DLG systems differ only slightly. Nevertheless, we believe that the effect can be observed in a wide range of $W$, simply because the Josephson current shows a qualitatively different behavior than the normal current. For $W=0$, the only current that may appear is the dc Josephson current in the case that the phases differ. Any finite gate voltage, on the other hand, introduces an ac Josephson current, while the current of unbound electrons is a dc current.

To conclude, the setup proposed might be used to identify a condensed exciton phase in DLG, which is subjected to an external electric potential, by analyzing a Josephson-type effect in a DLG junction device.  Provided the gate potentials of the DLG systems differ, an ac Josephson tunnel  
is observed irrespective of the phase relationship of the excitonic order parameters in both subsystems. If both DLG subsystems are exposed to the same gate voltages but there is a finite phase difference between their exciton order parameters,  a dc current appears. Such a finite phase difference suggests a degeneracy of the ground state, i.e., a $U(1)$ symmetry. This symmetry is closely related to off-diagonal long range order and only in this situation 
the exciting insulator does represent a genuine exciton condensate~\cite{ZFB14}.

We like to emphasize  that small leakage currents, which may arise in an actual experiment, are linked to interlayer hopping of electrons and holes or interlayer exchange terms due to the Coulomb interaction.  These terms pin the phase $\phi_i$ to a specific value and therefore destroy the $U(1)$ symmetry~\cite{LY76a, LZ96}. Hence exciton condensation---in a strict sense---cannot occur and the excitonic insulator, if present, features a charge-density-wave state. Interlayer hopping and exchange terms such as $H_U \propto a_{\k+\q i}^\dagger a_{\k i}^\pdagger a_{\k'-\q i}^\dagger b_{\k' i}^\pdagger$, moreover, enforce a finite $\Delta_{\k i}$ at all temperatures and therefore prevent a true phase transition~\cite{LZ96}.

The present study should be considered as a first step towards a theoretical modeling of the excitonic Josephson effect. The mean-field approach used is certainly a crude approximation and any future (more detailed) analysis should rely on more elaborated methods that take into account
fluctuation and correlation effects. Also the treatment of the tunnel junction could be improved, e.g., by using a more realistic (material specific)  tunnel matrix element and including pair-breaking effects within the barrier.  
In double bilayer graphene, respectively, double few-layer graphene systems the mobility of the electrons and holes is significantly reduced compared with double-layer graphene. This reduction strengthens electron-hole correlations and corroborates an EI formation at relatively high temperatures~\cite{PNH13,ZPNP14}. For these systems the theory presented above has to be adapted, but the basic idea and the scenario worked out should remain valid. Work along these lines will definitely improve our understanding of the fascinating exciton condensation phenomenon.  

%\acknowledgments
This work was supported by the DFG through the special research program SFB 652, project B5.

\bibliography{ref} 
\bibliographystyle{apsrev}

\end{document}